# Triangle for the entropic index q of non-extensive statistical mechanics observed by Voyager 1 in the distant heliosphere

**L. F. Burlaga and A. F.-Viñas**

Laboratory for Solar and Space Physics, Code 612.2,
NASA Goddard Space Flight Center, Greenbelt, Maryland, 20771 USA
Telephone: (301) 286-5956   Fax: (301) 286-1433
E-mail: Leonard.F.Burlaga@nasa.gov

**Abstract**

Tsallis [1] identified a set of numbers, the "q-triplet" $\equiv \{q_{stat}, q_{sen}, q_{rel}\}$, for a system described by nonextensive statistical mechanics. The deviation of the q's from unity is a measure of the departure from thermodynamic equilibrium. We present observations of the q-triplets derived from two sets of daily averages of the magnetic field strength B observed by Voyager 1 in the solar wind near 40 A.U. during 1989 and near 85 A.U. during 2002, respectively. The results for 1989 do not differ significantly from those for 2002. We find $q_{stat} = 1.75 \pm 0.06$, $q_{sen} = -0.6 \pm 0.2$, and $q_{rel} = 3.8 \pm 0.3$.

# 1. Introduction.

Tsallis [1] discussed a generalization of Boltzmann-Gibbs (B-G) statistical mechanics, a nonextensive statistical mechanics, which has applications to many complex physical systems that are not in thermal equilibrium [2], [3] and [4]. A principal result [5] is the derivation of a probability distribution function (PDF), a "q-exponential function" $\exp_q[\beta x] \equiv [1 + (1 - q) \beta x]^{1/(1 - q)}$, from a new entropy principle and two constraints. The entropic parameter q in the PDF describes the nonextensivity of the system. When q = 1, the PDF of Tsallis statistical mechanics reduces to the Boltzmann-Gibbs distribution.

Boltzmann-Gibbs statistical mechanics has the following characteristics [1]. (1) The B-G PDF is an exponential function of energy describing a thermal equilibrium state characterized by the temperature T. (2) B-G statistics is related to exponential sensitivity to the initial conditions (strong chaos, described by an exponential growth characterized by one or more positive Lyapunov exponents). (3) B-G statistics is associated with the exponential relaxation of macroscopic quantities to thermal equilibrium (exponential decay with a relaxation time τ). Tsallis [1] suggested that, by analogy with B-G statistical mechanics, nonextensive statistical mechanics has the following characteristics. (1) The PDF is a q-exponential distribution describing a metastable or quasi-stationary state with the parameter q ≡ $q_{stat}$. (2) Such a system is related to q-exponential sensitivity to the initial conditions (weak chaos, described by a q-exponential growth characterized by zero Lyapunov exponents and a growth parameter $q_{sen}$.). (3) Tsallis statistics is associated with the q-exponential relaxation of macroscopic quantities to thermal equilibrium (exponential decay with a relaxation parameter $q_{rel}$). In general, the metastable state is characterized by a "q-triplet", $\{q_{stat}, q_{sen}, q_{rel}\} \neq \{1, 1, 1\}$ where $q_{stat} > 1$, $q_{sen} < 1$, and $q_{rel} > 1$. Ultimately, the system might relax to a state with $\{q_{stat}, q_{sen}, q_{rel}\} = \{1, 1, 1\}$, corresponding to the B-G thermal equilibrium state. To the best of our knowledge, observations of the q-triplet for a physical system have not yet been published.

The atmosphere of the Sun beyond a few solar radii (the heliosphere) is fully ionized plasma expanding at supersonic speeds, carrying solar magnetic fields with it [6] and [7]. This "solar wind" is a driven nonlinear non-equilibrium system. The Sun injects matter, momentum, energy and magnetic fields into the heliosphere in a highly variable way. At the orbit of Earth (1 A.U.), the fluctuations in speed V and magnetic field strength B are large. As a result of nonlinear interactions, the fluctuations in B grow in size and amplitude with increasing distance from the Sun out to a distance of 5 – 30 A.U. [8]. MHD simulations [9] and [10] show that between ≈40 A.U. and ≈90 A.U. (the "distant heliosphere"), the fluctuations can relax slowly, and the solar wind can be in a quasi-stationary, metastable state.

At 1 A.U., the fluctuations in B have a symmetric Tsallis distribution on scales from 1 hour to 128 days [11]. These results indicate that nonextensive statistical mechanics is useful for describing the solar wind and states of the heliosphere. Fluctuations in V and B in the solar wind, often have fractal and multifractal scaling structure over a large range of scales in the region from 1 – 85 A.U. [12] and [8]. This type of scaling represents a hierarchical structure in phase space, in contrast to the uniformly occupied phase space of Boltzmann-Gibbs (B-G) statistical mechanics. The multifractal spectrum of the fluctuations in B observed near 85 A.U. during 2002 is very similar to that observed near 40 A.U. during 1989 [12] and [13] suggesting the existence of a quasi-stationary, metastable state organized about a multifractal attractor in the phase space of the distant heliosphere, far from the driving source.

The aim of this paper is to determine the q-triplet for the solar wind in the distant heliosphere and to compare the properties of this q-triplet with those expected for a quasi-stationary metastable dynamical system described by nonextensive statistical mechanics.

## 2. Analysis and results

The Voyager 1 (V1) observations of daily averages of B for 1989 and 2002 made between 36.3 to 38.8 A.U. (≈40 A.U.) and between 83.4 to 86.9 A.U. (≈85 A.U.), respectively, are shown in Fig. 1 and Fig. 2, respectively. Two features of these time series are of particular significance: the presence of large amplitude fluctuations in B on a wide range of scales, and the occurrence of many large abrupt jumps in B.

The value of $q_{stat}$ is derived from a PDF. The large multiscale fluctuations in B can be described by the PDFs of $dBn(t_i) \equiv B(t_i + \tau_n) - B(t_i)$ on scales $\tau_n = 2^n$ days, where n = 0, 1, 2, 3, 4, 5, 6, and 7, i.e., scales from 1 to 128 days. The observed PDFs for the 1989 and 2002 are shown by the symbols in Fig. 2 and Fig. 3, respectively. The PDFs in each figure are displaced vertically relative to one another by a factor of 100 for clarity. The solid curves are fits of the observed PDFs to the Tsallis q-exponential distribution,

$$A_q \times \exp_q[-\beta_q \, (dBn)^2] \equiv A_q \times [1 - (1 - q) \, \beta_q \, (dBn)^2]^{1/(1-q)} \qquad (1)$$

where $q \equiv q_{stat}$ is the "entropic" or "nonextensivity" factor, $-\infty < q_{stat} \leq 3$ [5] and [11]. The Tsallis q-exponential distribution fits all of the observed PDFs in Fig. 2 and Fig. 3 very well, on scales from 1 to 128 days, at heliospheric distances of 40 and 85 A.U., and during the years 1989 and 2002. This result supports the hypothesis that the distant heliosphere might be in a state described by the nonextensive statistical mechanics of Tsallis.

The PDFs in Fig. 2 and Fig. 3 range from kurtotic distributions at small scales (with large tails related to jumps in B) toward a B-G distribution at the largest scales (which would appear as a parabolic curve on the semilog plots in Fig. 2 and Fig. 3). From a fit of the Tsallis distribution to the observations with $\tau = 1$ day we obtain $q_{stat} = 1.72 \pm 0.06$ and $q_{stat} = 1.77 \pm 0.06$ for the 1989 and 2002 observations, respectively. The two values of $q_{stat}$ are the same within the uncertainties, but one cannot exclude the possibility of a small systematic change from $\approx$40 A.U. to $\approx$85 A.U. As a representative value we take $q_{stat} = 1.75 \pm 0.06$.

The value of $q_{sen}$ can be derived from the multifractal spectrum $f(\alpha)$ obtained from B(t). The sensitivity to initial conditions of one-dimensional nonlinear maps is described by the q-exponential distribution with $q = q_{sen}$ (rather than the exponential distribution for strong chaos and mixing), which implies a power law sensitivity when $q_{sen} \neq 1$. The multifractal structure of the critical attractor is given by the multifractal spectrum $f(\alpha)$, whose end points $\alpha_{min}$ and $\alpha_{max}$ (where $f(\alpha) = 0$) are the singularity strengths where the measure is most (least) concentrated, respectively. A relation between $q_{sen}$ and multifractality in dissipative systems [14] is

$$1/(1 - q_{sen}) = 1/\alpha_{min} - 1/\alpha_{max} \qquad (2)$$

The daily averages of B observed by V1 during 2002 (1989) have a multifractal structure in the range of scales from $\approx$2 to 16 days ($\approx$2 to 32 days) [12, 13]. The measured values of ($\alpha_i$, $f_i$) for 1989 and 2002 are reproduced in Fig. 4(a) and Fig. 4(b), respectively. In order to determine $\alpha_{min}$ and $\alpha_{max}$ it is necessary to fit the observations with some function and extrapolate to $f(\alpha) = 0$. The uncertainties in $\alpha_{min}$ and $\alpha_{max}$ determine the uncertainty in $q_{sen}$, but these uncertainties are largely determined by the choice of the function $f(\alpha)$ used to fit the data. The theoretical form of $f(\alpha)$ is not known, although one expects it to be a concave function with a single maximum [15]. A quadratic function, shown by the curves in Fig. 4(a) and Fig. 4(b), provides a good fit to the observations, but the fit is not unique. A cubic fit also provides a good fit to both sets of the observations over the observed range of alpha, but extrapolation of the cubic gives an unphysical inflection point for $\alpha > 1$. Two quadratic functions, one for $\alpha > 1$ and another for $\alpha < 1$, are also consistent with each of the data sets. Based on these and other fitting functions we estimate the following: 1) $f_{max}$= 1.02 ± 0.06 at $\alpha$ = 1.02 ± 0.02; 2) $\alpha_{min}$ = 0.70 ± 0.06 for the 1989 data and 0.76 ± 0.06 for the 2002 data; and 3) $\alpha_{max}$ = 1.4 ± 0.1 for the 1989 data and 1.4 ± 0.1 for the 2002 data. Thus, the observations of $\alpha_{min}$ and $\alpha_{max}$ for the 1989 data are the same as those for the 2002 data, within the uncertainties, consistent with a quasi-stationary metastable state of the heliosphere between 40 and 85 A.U. The quadratic fit in Fig. 4(a) is broader than that in Fig. 4(b), but this difference is not significant within the uncertainties. From (2) one obtains $q_{sen}$ = -0.4 ± 0.2 for 1989, $q_{sen}$ = -0.7 ± 0.2 for 2002. We take as a representative value for the distant heliosphere $q_{sen}$ = -0.6 ± 0.2.

The value of $q_{rel}$ can be determined from a scale-dependent correlation coefficient $C(\tau)$ defined as follows:

$$C(\tau) \equiv <[B(t_i + \tau) - <B(t_i)>] \times [B(t_i) - <B(t_i)>)]>/<[(B(t_i) - <B(t_i)>]^2> \qquad (3)$$

In general, one expects $C(\tau)$ to evolve with distance from the Sun and to be a function of the solar cycle. In classical B-G statistical processes this correlation coefficient tends to decrease exponentially with increasing $\tau$, but for Tsallis statistics it should decay as a power law for some range of lags. In the latter case, log $C(\tau)$ = a + s log $\tau$, where the slope s = 1/(1 - $q_{rel}$), and $q_{rel}$ describes a relaxation process [16]. We find that $C(\tau)$ does decreases exponentially with increasing $\tau$ near the Sun, at 1 A.U.. However, at ≈40 A.U. (1989) and ≈85 A.U. (2002), $C(\tau)$ decreases as a power law with increasing $\tau$ over a certain range, as shown in Fig. 5(a) and Fig. 5(b), respectively. The solid lines in Fig. 5(a) and Fig. 5(b) show a power law decay on scales from 1 to 8 days, giving $q_{rel}$ = 4.1 and 3.5 for 1989 and 2002, respectively. A straight line also provides relatively good fits to the data on scales from 1 to 16 days, giving $q_{rel}$ = 3.9 and 3.5 for 1989 and 2002, respectively. The uncertainties in $q_{rel}$ depend primarily on the range over which one fits the observations. As a representative value of $q_{rel}$ in the distant heliosphere we take $q_{rel}$ = 3.8 ± 0.3.

## 3. Conclusions

In summary, using in situ measurements in the distant heliosphere we calculated the q-triplet of nonextensive statistical mechanics $\{q_{stat}, q_{sen}, q_{rel}\}$. We estimate all three "q"s from the magnetic field strength observations made by V1 at $\approx$ 40 A.U. during 1989 and at $\approx$ 85 A.U. during 2002. The same results (within the uncertainties) were obtained for both sets of data, supporting the hypothesis that the distant heliosphere tends to be in a quasi-stationary metastable state. We found that $q_{stat} = 1.7 \pm 0.06$, $q_{sen} = -0.6 \pm 0.2$, and $q_{rel} = 3.8 \pm 0.3$.

**Acknowledgments**

We thank Prof. Tsallis for introducing us to the q-triplet and for stimulating discussions. The data in this paper are from the magnetic field experiment on Voyager 1. The Principal Investigator is N. F. Ness. We thank M. Acuña for discussions concerning the instrument. T. McClanahan and S. Kramer carried out the processing of the data.

Fig. 1. (a) Daily averages of the magnetic field strength B versus time measured by Voyager 1 (V1) during 1989. (b) Daily averages of B versus time measured by V1 during 2002.

Fig. 2. The symbols are PDFs of relative changes in the magnetic field strength measured by V1 during 1989 on scales from 1 to 128 days. The curves are fits of the data to the q-exponential distribution function.

Fig. 3. The symbols are PDFs of relative changes in the magnetic field strength measured by V1 during 2002 on scales from 1 to 128 days. The curves are fits of the data to the q-exponential distribution function.

Fig. 4. (a) The symbols are based on measurements of the multifractal spectrum f($\alpha$) versus $\alpha$ determined from magnetic field strength observations made by V1 during 1989. (b) The symbols are based on measurements of the multifractal spectrum f($\alpha$) versus $\alpha$ determined from magnetic field strength observations made by V1 during 2002. The curves in (a) and (b) are quadratic fits to the data, one of several types of curves used to estimate $\alpha_{min}$ and $\alpha_{max}$.

Fig. 5. (a) The correlation coefficient C($\tau$) versus scale $\tau$ computed from daily averages of B measured by V1 during 1989. (b) The correlation coefficient C($\tau$) versus scale $\tau$ computed from daily averages of B measured by V1 during 2002. The lines in (a) and (b) are linear least squares fits to the data in the range $\tau = 1 - 8$ days.

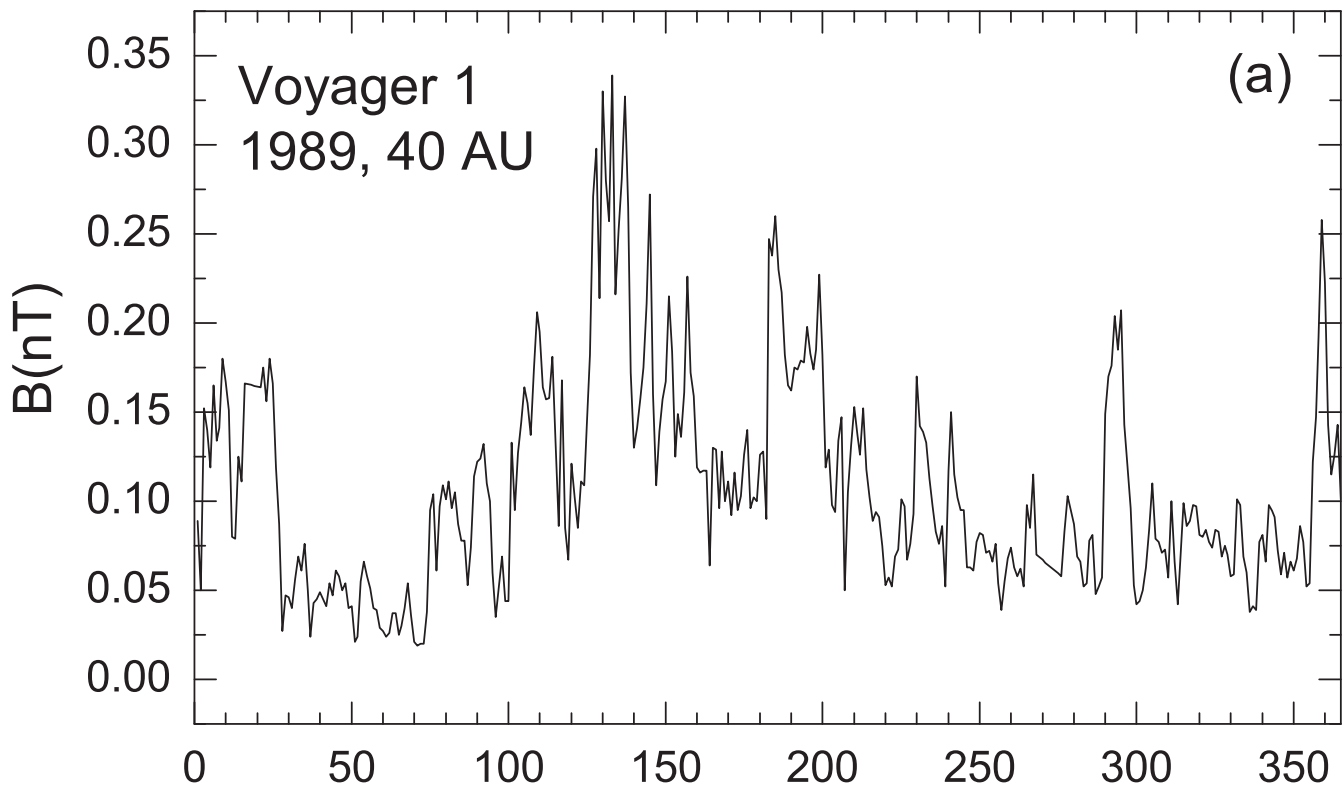
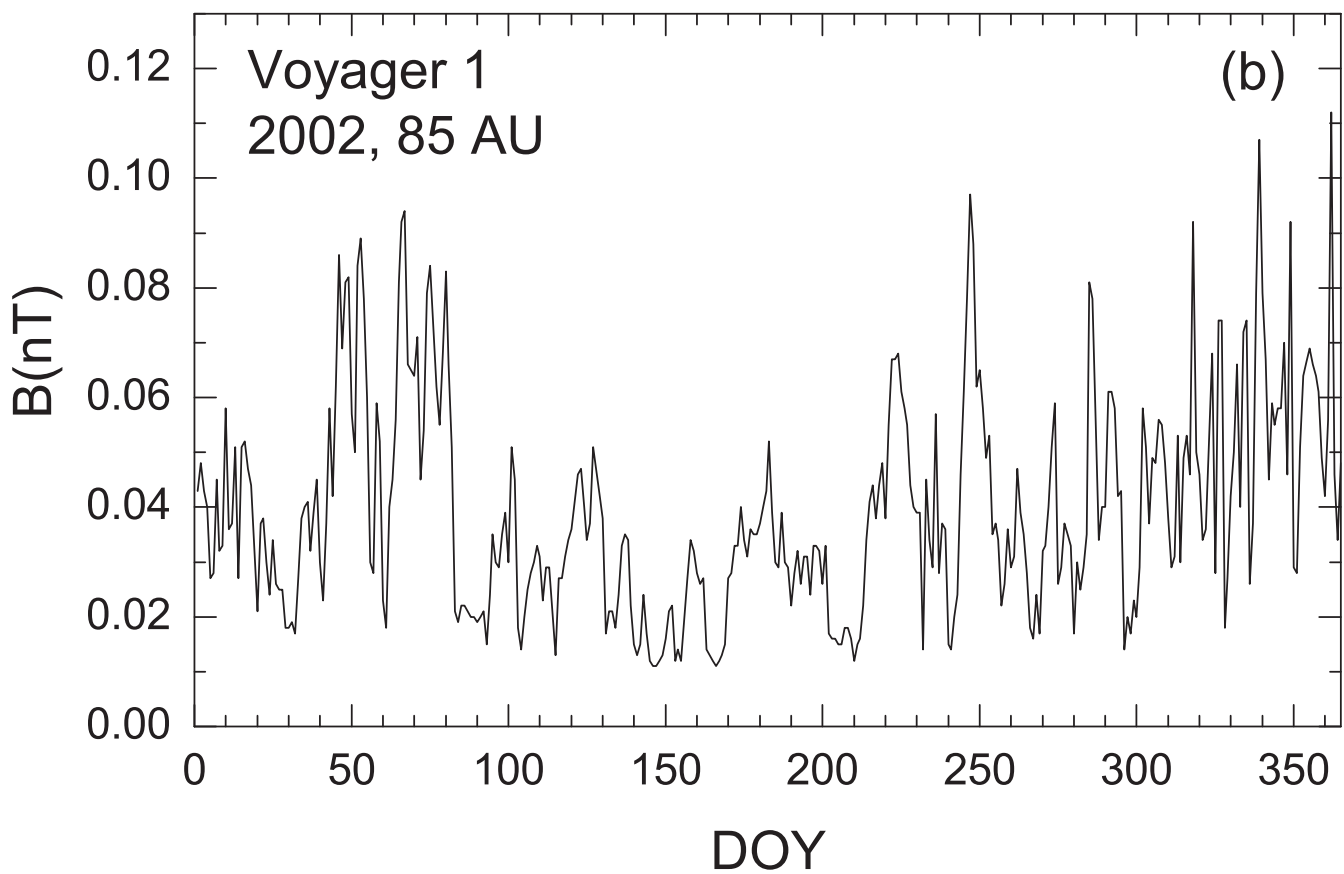

Fig. 1. Burlaga & Vinas

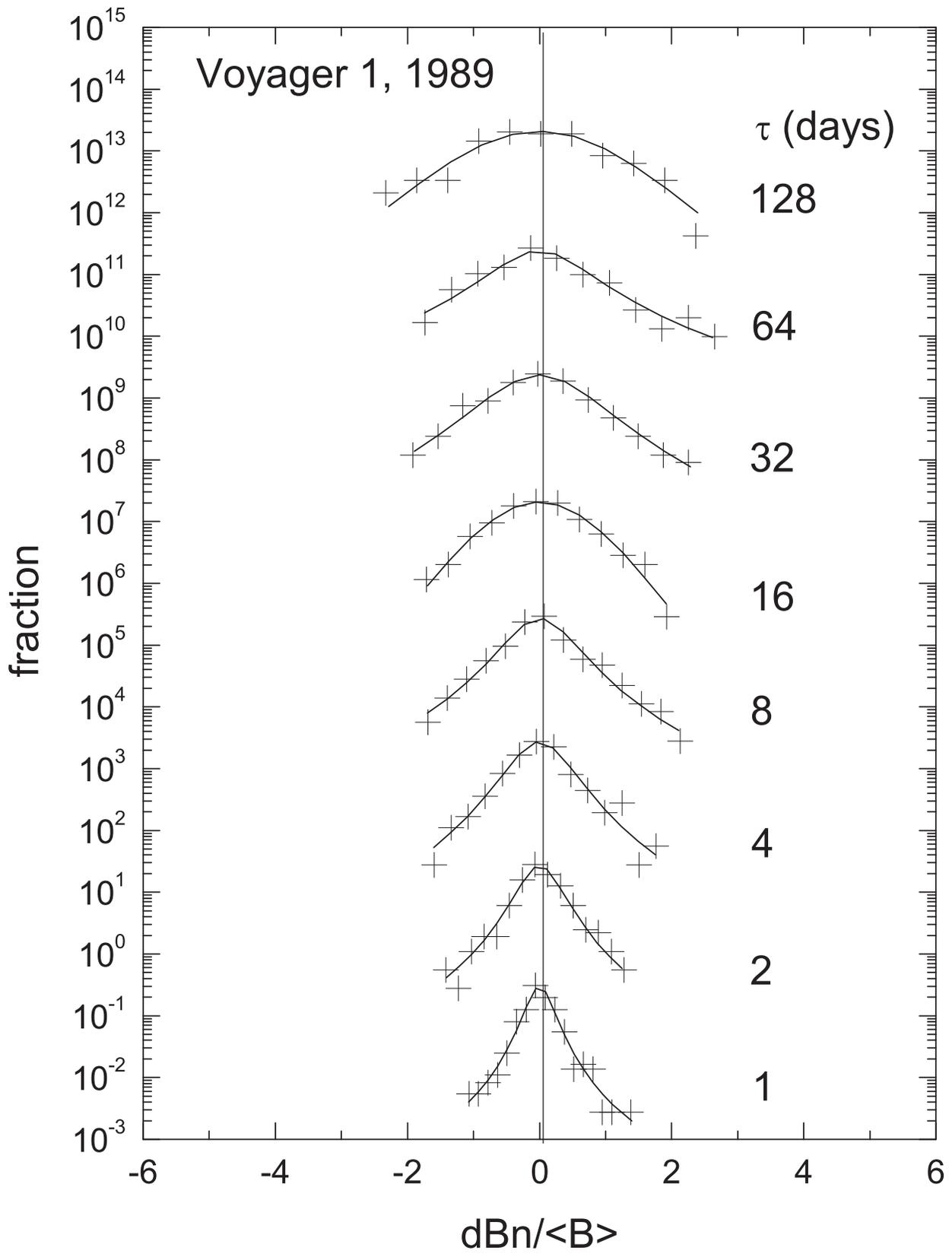

Fig. 2. Burlaga & Vinas

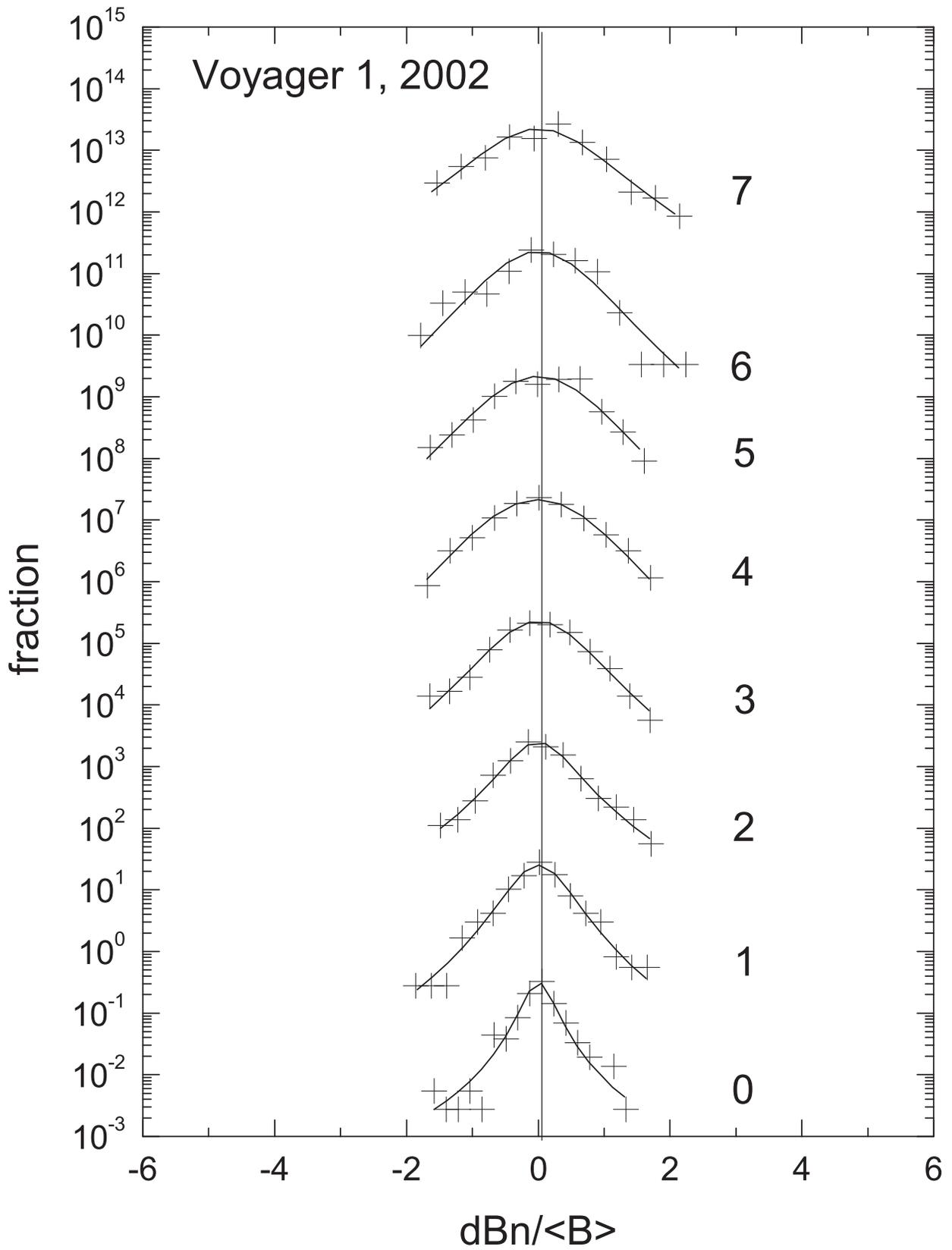

Fig. 3. Burlaga & Vinas

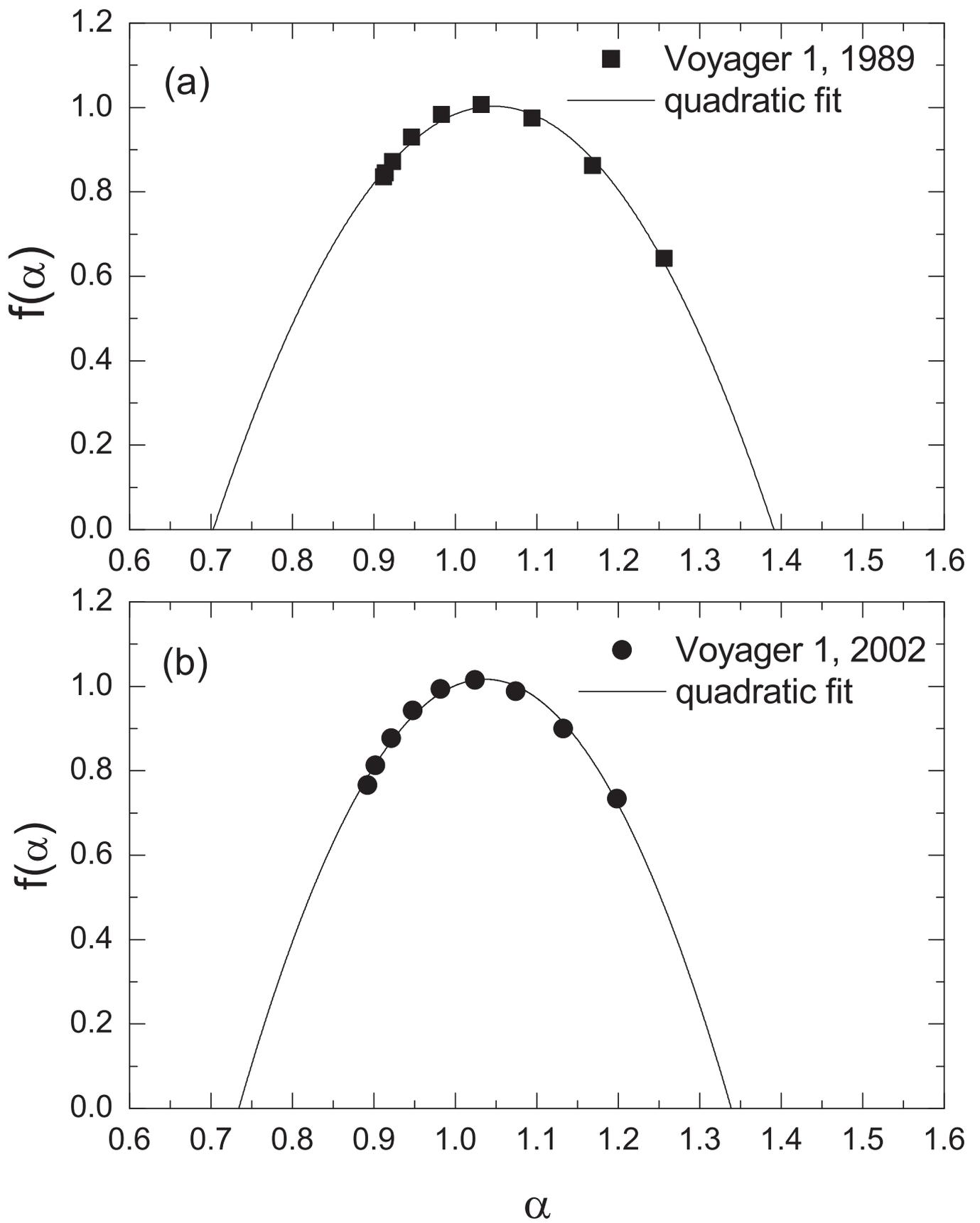

Fig. 4. Burlaga & Vinas

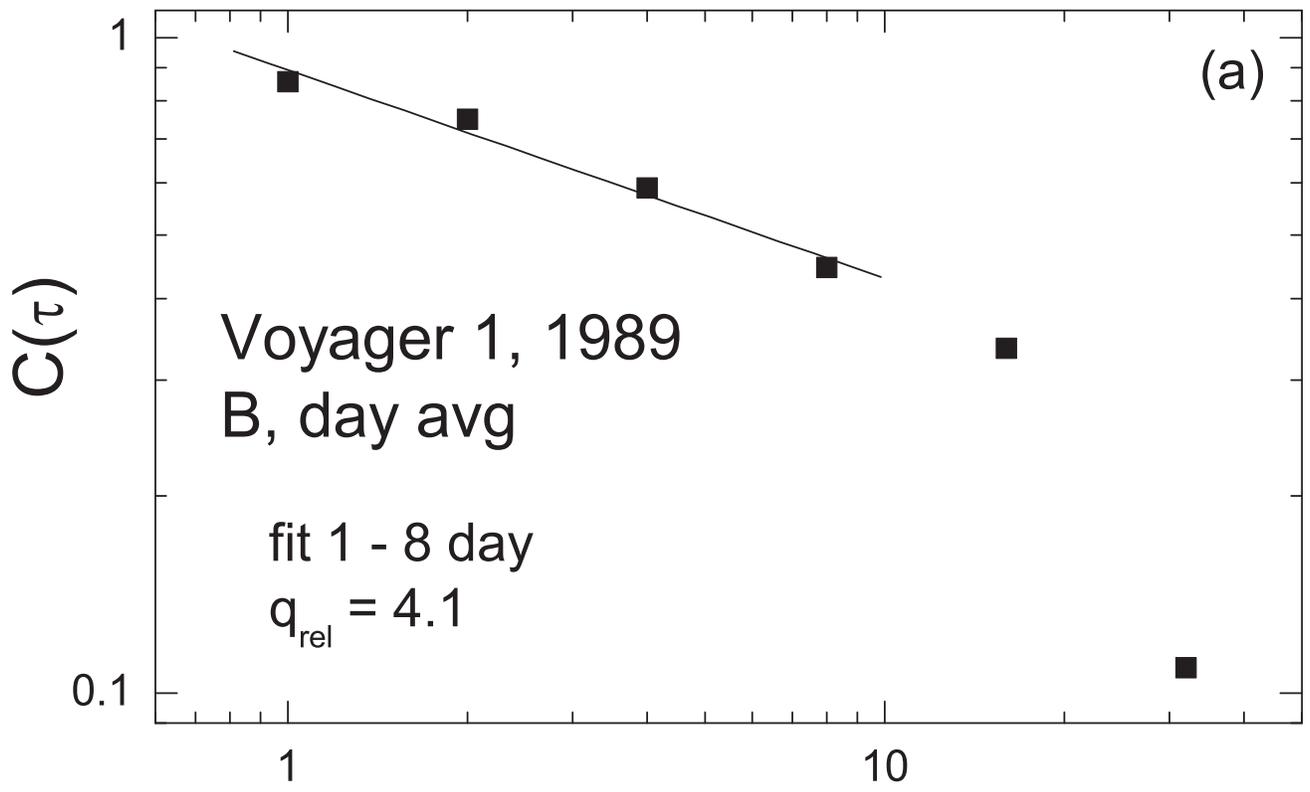
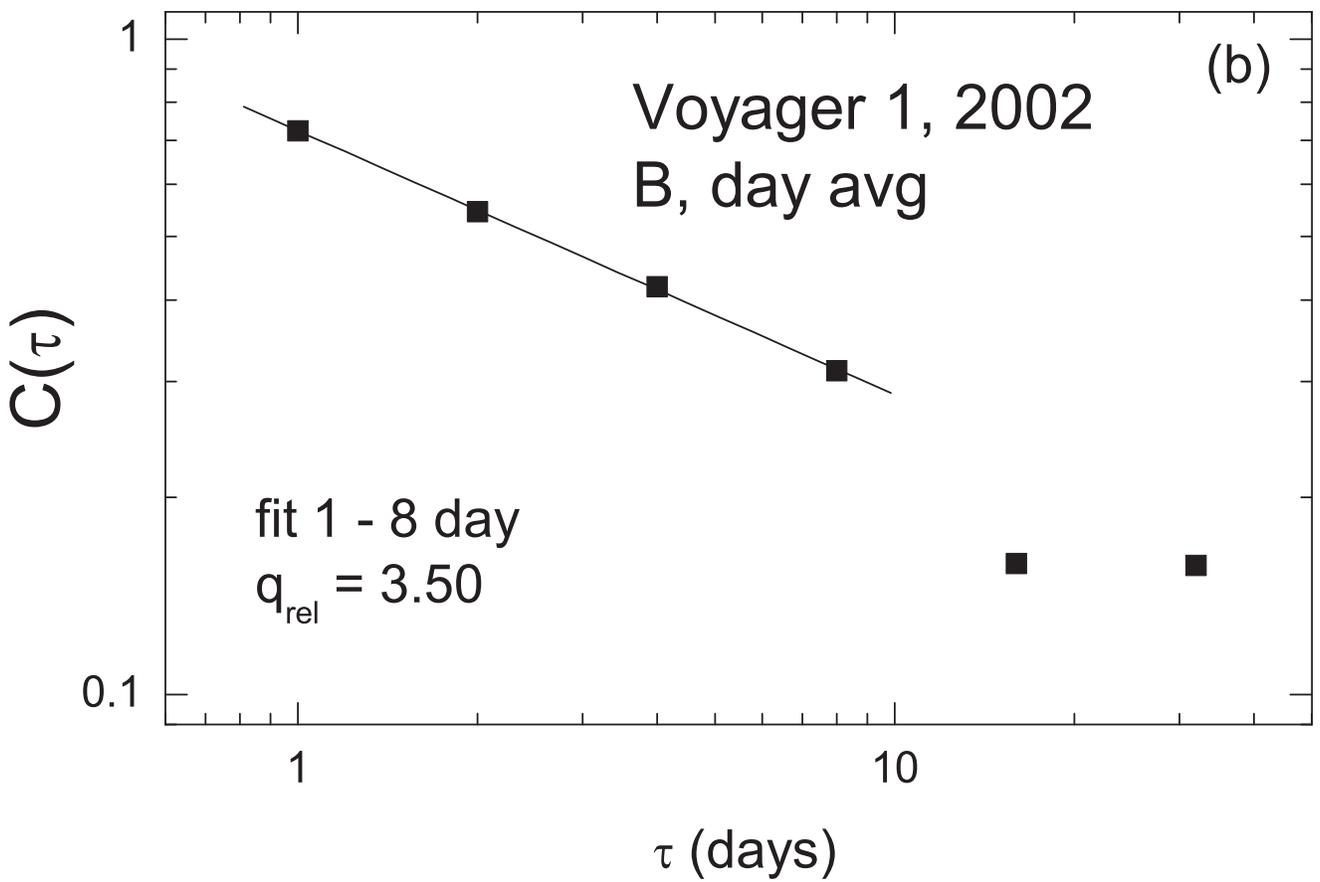

Fig. 5. Burlaga & Vinas